\documentclass[epj]{svjour}
\usepackage{graphicx}
\usepackage{amssymb}

\begin{document}

\title{Magnetic properties and temperature variation of spectra in the Hubbard model}%
\author{Alexei Sherman}%
\institute{Institute of Physics, University of Tartu, W. Ostwaldi Str 1, 50411 Tartu, Estonia\\
\email{alekseis@ut.ee}}

\date{Received: date / Revised version: date}

\abstract{
In the two-dimensional fermionic Hubbard model, temperature and concentration dependencies of the uniform magnetic susceptibility and squared site spin, the variation of the double occupancy with the repulsion and the temperature dependence of the spin structure factor are calculated using the strong coupling diagram technique. In these calculations, a correction parameter is introduced into the irreducible vertex to fulfill the Mermin-Wagner theorem and to attain low temperatures. Satisfactory agreement of the obtained results with data of Monte Carlo simulations, numerical linked-cluster expansions and experiments in optical lattices lends support to the validity of such a correction. The ability to attain low temperatures allows us to investigate spectral functions in this region. At half-filling, for small and large Hubbard repulsions no qualitative changes are observed in comparison with somewhat higher temperatures reached in the previous work. However, on cooling, there appears a new feature for moderate repulsions –- a narrow band emerges near the Fermi level, which produces a pronounced peak in the density of states. By its location and bandwidth, the feature is identified with the spin-polaron band.
\PACS{{xx.xx.xx}{xx}}
}

\maketitle

\section{Introduction}
The repulsive two-dimensional (2D) Hubbard model is widely used for investigating strong electron correlations in cuprate perovskites. For its treatment, together with the numerical methods -- Monte Carlo simulations \cite{Hirsch,Moreo,Grober} and exact diagonalization \cite{Lin,Dagotto,Ebrahimkhas} -- the dynamic mean-field theory (DMFT) \cite{Georges} and its generalizations are used. Of these ge\-ne\-ra\-li\-za\-ti\-ons the cellular DMFT \cite{Maier,Park,Sato}, dynamic cluster approximation \cite{Maier,Moukouri,Merino14}, dynamic vertex approximation \cite{Toschi,Schafer,Rohringer} and dual fermion approach \cite{Rohringer,Rubtsov08,Hafermann} can be mentioned. Besides, there are several methods, which are not based on the DMFT. These are the cluster perturbation theory \cite{Senechal00,Senechal04,Kohno}, variation cluster approximation \cite{Potthoff,Arrigoni,Faye}, diagram technique for Hubbard operators \cite{Zaitsev,Izyumov88,Izyumov90,Ovchinnikov}, non-perturbative many-body approach \cite{Vilk}, numerical linked-cluster expansions (NLCE) \cite{Khatami,Tang}, fi\-ni\-te-tem\-pe\-ra\-tu\-re Lanczos method \cite{Jaklic,Bonca} and strong coupling diagram technique (SCDT) \cite{Vladimir,Metzner,Pairault,Sherman06,Sherman15,Sherman18}.

In this work, the SCDT is used for investigating spectral and magnetic properties of the model. The approach applies series expansions in powers of the kinetic energy for calculating Green's functions. In doing so infinite sequences of diagrams describing interactions of electrons with spin and charge fluctuations are taken into account. Obtained expressions form a closed set of equations, which can be self-consistently solved. Spectral functions and densities of states (DOS) de\-ri\-ved in this way were shown to be in satisfactory agreement with Monte Carlo and exact-diagonalization data as well as results of some other approaches for different sets of parameters \cite{Sherman18}. In comparison with these approaches, the SCDT has a number of advantages. Its numeric algorithm is simpler and requirements for the computation technique are very modest. The SCDT is an analytic method, which simplifies the interpretation of obtained results. In contrast to approaches based on the DMFT, all calculations are performed for an actual dimensionality of the model. As opposed to approaches using joining of cluster solutions or cluster generalizations of DMFT the used method does not reduce the original translation symmetry of the problem and allows one to consider charge and spin fluctuations of all ranges. Conceptually, the SCDT is close to the diagram technique for Hubbard operators. However, the number of diagrams describing the same processes is much smaller in SCDT. An additional difficulty of the technique for Hubbard operators is the dependence of its rules and graphic representation on the choice of the operator precedence. In the SCDT this problem is absent.

At the same time, the SCDT has a number of drawbacks. One of them is the violation of the Mermin-Wagner theorem \cite{Mermin}. The approach is able to describe the transition to the long-range antiferromagnetic (AF) order; however, the value of the transition temperature $T_{\rm AF}$ is finite for a 2D infinite half-filled crystal for the set of diagrams used in \cite{Sherman18}. In that paper, a way was pointed to remedy this defect. The improvement consists in introducing a parameter $\zeta$ in the irreducible vertex, which takes into account corrections containing higher-order cumulants. Due to the complexity of these corrections, it is difficult to calculate this parameter from the series expansion. In the SCDT, the AF transition is related to the vanishing determinant of a system of linear equations, and its zero value at $T=0$ is a natural condition for determining the parameter. This way is used in the present work that allows us to shift $T_{\rm AF}$ to zero and to consider the region of low temperatures. Another possible drawback of the SCDT is connected with the used expansion in powers of the kinetic energy. It can be expected that such an expansion is well suited for strong repulsions and works worse for small $U$. However, the mentioned self-consistency of the approach and the due regard for charge fluctuations seem to eliminate this defect -- for a repulsion as small as $U\approx t$ spectral functions have shapes inherent in the Slater regime \cite{Slater} and the double occupancy differs only by a few percents from the value obtained in Monte Carlo simulations (see below).

In this work, we test the applicability of the SCDT for the description of magnetic properties, verify the validity of the introduced correction and use the emerged possibility to attain lower temperatures for investigating spectral functions in this region. For these purposes, we calculate the temperature and concentration dependencies of the uniform magnetic susceptibility and the square of the site spin, the variation of the double occupancy with $U$ and temperature dependence of the spin structure factor in wide ranges of parameters and compare obtained results with data of Monte Carlo simulations, numerical linked-cluster expansions and experiments with ultracold fermionic atoms in 2D optical lattices. They are in satisfactory agreement up to temperatures as low as $T\approx 0.1t$ and repulsions as small as $U=t$. This lends support to the validity of the correction. For half-filling and low temperatures, we find no qualitative changes in spectra for $U\gtrsim 7t$ and $U\lesssim 3t$ in comparison with results obtained in the previous work \cite{Sherman18} at somewhat higher temperatures -- the former case is characterized by the four-band structure with the Mott gap near the Fermi level and two reabsorption pseudogaps around $\pm U/2$, while the latter case by two bands separated by the Slater gap along the boundary of the magnetic Brillouin zone. Qualitative changes occur for moderate repulsions $4t\lesssim U\lesssim 6t$, for which at low $T$ there appears a narrow band near the Fermi level. Its bandwidth of the order of the exchange constant $J=4t^2/U$ points to its spin-polaron character -- the band is formed from bound states of electrons and spin excitations. In this respect the band is similar to its counterparts observed for stronger correlations in the $t$-$J$ \cite{Schmitt,Ramsak,Sherman94} and Hubbard \cite{Sherman18} models with doping. The peak in the DOS stemming from the spin-polaron band is similar to the quasiparticle resonance of the DMFT and possesses conformable properties -- the peak is located at the FL in a wide range of electron concentrations and, for half-filling, it is observed for $U<U_c$, where $U_c\approx 6t$ is the critical repulsion of the Mott metal-insulator transition. The difference between the DMFT and SCDT peaks is in the character of quasiparticles forming related bound states. The DMFT peak is a modified Abrikosov-Suhl resonance, which originates from bound states of {\em free} electrons and {\em localized} spins. In the SCDT, the peak is a manifestation of bound states of {\em correlated} electrons and {\em mobile} spin excitations.

\section{Model and SCDT method}
The Hamiltonian of the 2D fermionic Hubbard model \cite{Hubbard63,Hubbard64} reads
\begin{equation}\label{Hamiltonian}
H=\sum_{\bf ll'\sigma}t_{\bf ll'}a^\dagger_{\bf l'\sigma}a_{\bf l\sigma}
+\frac{U}{2}\sum_{\bf l\sigma}n_{\bf l\sigma}n_{\bf l,-\sigma},
\end{equation}
where 2D vectors ${\bf l}$ and ${\bf l'}$ label sites of a square plane lattice, $\sigma=\pm 1$ is the spin projection, $a^\dagger_{\bf l\sigma}$ and $a_{\bf l\sigma}$ are electron creation and annihilation operators, $t_{\bf ll'}$ is the hopping constant and $n_{\bf l\sigma}=a^\dagger_{\bf l\sigma}a_{\bf l\sigma}$. In this work, only the nearest neighbor hopping constant $t$ is taken to be nonzero.

The following one- and two-particle Green's functions
\begin{eqnarray}
G({\bf l',\tau';l,\tau})&=&\langle{\cal T}\bar{a}_{\bf l'\sigma}(\tau')
a_{\bf l\sigma}(\tau)\rangle,\label{Glt}\\
\chi({\bf l',\tau';l,\tau})&=&\langle{\cal T}s^+_{\bf l'}(\tau')
s^-_{\bf l}(\tau)\rangle\nonumber\\
&=&\langle{\cal T}\bar{a}_{\bf l'\uparrow}(\tau')a_{\bf l'\downarrow}(\tau')\bar{a}_{\bf l\downarrow}(\tau)a_{\bf l\uparrow}(\tau)\rangle\label{clt}
\end{eqnarray}
are considered. Here the statistical averaging denoted by the angular brackets and time dependencies $$\bar{a}_{\bf l\sigma}(\tau)=\exp{({\cal H}\tau)}a^\dagger_{\bf l\sigma}\exp{(-{\cal H}\tau)}$$
are determined by the operator ${\cal H}=H-\mu\sum_{\bf l\sigma}n_{\bf l\sigma}$ with the chemical potential $\mu$. The time-ordering operator ${\cal T}$ arranges operators from right to left in ascending order of times $\tau$. Up to the constant factor $\chi({\bf l',\tau';l,\tau})$ (\ref{clt}) coincides with the spin susceptibility.

\begin{figure}[t]
\centerline{\resizebox{0.95\columnwidth}{!}{\includegraphics{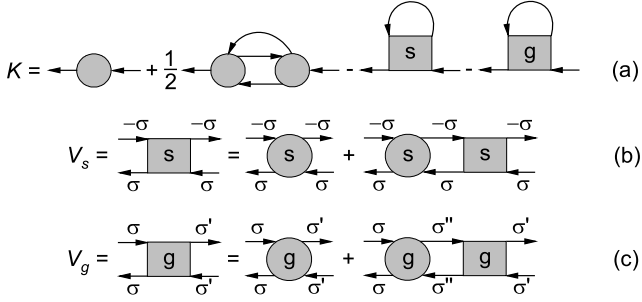}}}
\caption{(a) Diagrams in $K({\bf k},j)$ taken into account in the present work. (b,c) Bethe-Salpeter equations for the four-leg diagrams in part (a).} \label{Fig1}
\end{figure}

For calculating these quantities the SCDT \cite{Vladimir,Metzner,Pairault,Sherman06,Sherman15} is used. In this approach, Green's functions are represented by the series expansion in powers of $t_{\bf ll'}$, each term of which is a product of the hopping constants and on-site cumulants of creation and an\-ni\-hi\-la\-ti\-on operators. These terms can be visualized as a sequence of directed lines corresponding to the hopping constants $t_{\bf ll'}$, which connect circles picturing cumulants of different orders (a concise description of this diagram technique can be found in \cite{Sherman16,Sherman17}). For the one-particle Green's function (\ref{Glt}), all these terms can be summed, and after the Fourier transformation the result reads
\begin{equation}\label{Larkin}
G({\bf k},j)=\Big\{\big[K({\bf k},j)\big]^{-1}-t_{\bf k}\Big\}^{-1},
\end{equation}
where ${\bf k}$ is the 2D wave vector, $j$ is an integer defining the Matsubara frequency $\omega_j=(2j-1)\pi T$, $t_{\bf k}$ is the Fourier transform of $t_{\bf ll'}$ and $K({\bf k},j)$ is the irreducible part -- the sum of all two-leg irreducible diagrams, which cannot be divided into two disconnected parts by cutting a hopping line. Diagrams taken into account in this work for calculating $K({\bf k},j)$ are shown in figure \ref{Fig1}(a).
The linked-cluster theorem is valid and partial summations are allowed in the SCDT. Thanks to this possibility, bare internal lines $t_{\bf k}$ in this figure can be transformed into dressed ones,
\begin{equation}\label{hopping}
\theta({\bf k},j)=t_{\bf k}+t^2_{\bf k}G({\bf k},j).
\end{equation}

In figure \ref{Fig1}(a), circles without letters are on-cite cumulants of the first and second orders
\begin{eqnarray*}
&&C_1(\tau',\tau)=\big\langle{\cal T}\bar{a}_{{\bf l}\sigma}(\tau')a_{{\bf l}\sigma}(\tau)\big\rangle_0,\\
&&C_2(\tau_1,\sigma_1;\tau_2,\sigma_2;\tau_3,\sigma_3;\tau_4,\sigma_4)\\
&&\quad\quad=\big\langle{\cal T}\bar{a}_{{\bf l}\sigma_1}(\tau_1)a_{{\bf l}\sigma_2}(\tau_2) \bar{a}_{{\bf l}\sigma_3}(\tau_3)a_{{\bf l}\sigma_4}(\tau_4)\big\rangle_0\\
&&\quad\quad-\big\langle{\cal T}\bar{a}_{{\bf l}\sigma_1}(\tau_1)a_{{\bf l}\sigma_2}(\tau_2)\big\rangle\big\langle{\cal T}\bar{a}_{{\bf l}\sigma_3}(\tau_3)a_{{\bf l}\sigma_4}(\tau_4)\big\rangle_0\\
&&\quad\quad+\big\langle{\cal T}\bar{a}_{{\bf l}\sigma_1}(\tau_1)a_{{\bf l}\sigma_4}(\tau_4)\big\rangle\big\langle{\cal T}\bar{a}_{{\bf l}\sigma_3}(\tau_3)a_{{\bf l}\sigma_2}(\tau_2)\big\rangle_0,
\end{eqnarray*}
where the subscript 0 near brackets indicates that time dependencies and averages are determined by the local operator ${\cal H}_{\bf l}=\sum_\sigma\left[(U/2)n_{\bf l\sigma}n_{\bf l,-\sigma}- \mu n_{\bf l\sigma}\right]$. Due to the translation symmetry the cumulants are identical on all lattice sites. The last two terms in $K({\bf k},j)$ correspond to diagrams with ladder inserts. The vertices $V_s$ and $V_g$ satisfy the Bethe-Salpeter equations shown in figures~\ref{Fig1}(b) and (c), in which circles with letters are irreducible four-leg diagrams which cannot be divided into two disconnected parts by cutting a pair of horizontal oppositely directed hopping lines. In this work, these irreducible diagrams are approximated by the respective second-order cumulants. In this case, $V_s$ and $V_g$, apart from frequencies, depend only on the transfer momentum. Therefore, the irreducible part reads
\begin{eqnarray}\label{K}
&&K({\bf k},j)=C_1(j)\nonumber\\
&&\quad\quad-\frac{T}{N}\sum_{{\bf k'}j'}\theta({\bf k'},j')\big[V_{s,\bf k- k'}(j,\sigma;j,\sigma;j',-\sigma;j',-\sigma)\nonumber\\
&&\quad\quad+V_{g,\bf k-k'}(j,\sigma;j,\sigma;j',\sigma;j',\sigma)\big]\nonumber\\
&&\quad\quad+\frac{T^2}{2N^2}\sum_{{\bf k'}j'\nu}\theta({\bf k'},j'){\cal T}_{\bf k-k'}(j+\nu,j'+\nu) \nonumber\\
&&\quad\quad\times\Big[C_2(j,\sigma;j+\nu,\sigma;j'+\nu,-\sigma;j',-\sigma)\nonumber\\
&&\quad\quad\times C_2(j+\nu,\sigma;j,\sigma;j',-\sigma;j'+\nu,-\sigma)\nonumber\\
&&\quad\quad+\sum_{\sigma'} C_2(j,\sigma;j+\nu,\sigma';j'+\nu,\sigma';j',\sigma)\nonumber\\
&&\quad\quad\times C_2(j+\nu,\sigma';j,\sigma;j',\sigma;j'+\nu,\sigma')\Big].
\end{eqnarray}

The vertex $V_s$ is connected with the spin su\-s\-cep\-ti\-bi\-li\-ty (\ref{clt}) by the relation
\begin{eqnarray}\label{chi}
\chi({\bf k},\nu)&=&-\frac{T}{N}\sum_{{\bf q}j}G({\bf q},j)G({\bf q+k},j+\nu) \nonumber\\
&&-T^2\sum_{jj'}F_{\bf k}(j,j+\nu)F_{\bf k}(j',j'+\nu)\nonumber\\
&&\times V_{s\bf k}(j+\nu,\sigma;j'+\nu,\sigma;j',-\sigma;j,-\sigma),
\end{eqnarray}
where
\begin{eqnarray*}
&&F_{\bf k}(j,j')=N^{-1}\sum_{\bf q}\Pi({\bf q},j)\Pi({\bf q+k},j'),\\
&&\Pi({\bf k},j)=1+t_{\bf k}G({\bf k},j),
\end{eqnarray*}
$N$ is the number of sites and $\nu$ defines the boson Matsubara frequency $\omega_\nu=2\nu\pi T$. It can be shown that $V_s$ coincides with $\sigma\sum_{\sigma'}\sigma'V_g(\sigma',\sigma,\sigma,\sigma')$, the antisymmetrized part of $V_g$, while $V_c=\sum_{\sigma'}V_g(\sigma',\sigma,\sigma,\sigma')$, the sym\-met\-ri\-zed part is connected with the charge susceptibility by the relation similar to (\ref{chi}).

Equations describing second-order cumulants are given in \cite{Vladimir,Pairault,Sherman06,Sherman07,Sherman08}. These equations contain terms proportional to $1/T$. The divergence of these terms at $T\rightarrow 0$ leads to overestimating the interaction of electrons with spin fluctuations. In its turn, it results in a finite value of $T_{\rm AF}$, in contradiction with the Mermin-Wagner theorem. To remedy this defect, in the present work, the multiplier $1/T$ is substituted with $1/(T+\zeta)$, and the parameter $\zeta$ is determined from the condition $T_{\rm AF}=0$ at half-filling when the tendency toward the AF ordering is most pronounced. This procedure will be discussed in more detail below.

The equations for the second-order cumulants are ra\-ther cumbersome. However, they can be significantly simplified in the case
\begin{equation}\label{condition}
T\ll\mu,\quad T\ll U-\mu.
\end{equation}
For $U\gg T$ this range of $\mu$ contains the most interesting cases of half-filling, $\mu=U/2$, and moderate doping. For the conditions (\ref{condition}) and with the introduced parameter $\zeta$ the first- and second-order cumulants read
\begin{eqnarray}
&&C_1(j)=\frac{1}{2}\left[g_1(j)+g_2(j)\right],\nonumber\\
&&C_2(j+\nu,\sigma;j,\sigma';j',\sigma';j'+\nu,\sigma)\nonumber\\[-1.5ex]
&&\label{cumulants}\\[-1.5ex]
&&\quad=\frac{1}{4(T+\zeta)}\big[\delta_{jj'} \big(1-2 \delta_{\sigma\sigma'}\big)
+\delta_{\nu 0}\big(2-\delta_{\sigma\sigma'}\big)\big]\nonumber\\
&&\quad\times a_1(j'+\nu)a_1(j)-\delta_{\sigma,-\sigma'}B(j,j',\nu),\nonumber
\end{eqnarray}
where
\begin{eqnarray}
&&g_1(j)=(i\omega_j+\mu)^{-1},\quad g_2(j)=(i\omega_j+\mu-U)^{-1}, \nonumber\\
&&B(j,j',\nu)=\frac{1}{2}\big[a_1(j'+\nu)a_2(j,j') \nonumber\\
&&\;+a_2(j'+\nu,j+\nu)a_1(j)+a_4(j'+\nu,j+\nu)a_3(j,j') \nonumber\\[-0.5ex]
&&\label{terms}\\[-1.5ex]
&&\;+a_3(j'+\nu,j+\nu)a_4(j,j')\big], \nonumber\\
&&a_1(j)=g_1(j)-g_2(j),\quad a_2(j,j')=g_1(j)g_1(j'), \nonumber\\
&&a_3(j,j')=g_2(j)-g_1(j'),\quad a_4(j,j')=a_1(j)g_2(j'). \nonumber
\end{eqnarray}

With these expressions for the cumulants, the Bethe-Salpeter equations in figures~\ref{Fig1}(b) and (c) can be written as follows:
\begin{eqnarray}\label{Vs}
&&V_{s\bf k}(j+\nu,j,j',j'+\nu)=\frac{1}{2}f_{1\bf k}(j+\nu,j'+\nu)\nonumber\\
&&\quad\quad\times\bigg\{\bigg[a_2(j'+\nu,j+\nu)-\frac{\delta_{jj'}}{T+\zeta} a_1(j'+\nu)\bigg] \nonumber\\
&&\quad\quad\times\big[a_1(j)+y_{1\bf k}(j,j')\big]-\frac{\delta_{\nu0}}{2(T+\zeta)}a_1(j)a_1(j') \nonumber\\
&&\quad\quad+a_1(j'+\nu)\big[a_2(j,j')+y_{2\bf k}(j,j')\big]\nonumber\\
&&\quad\quad+a_3(j'+\nu,j+\nu)\big[a_4(j,j')+y_{4\bf k}(j,j')\big]\nonumber\\
&&\quad\quad+a_4(j'+\nu,j+\nu)\big[a_3(j,j')+y_{3\bf k}(j,j')\big]\bigg\},
\end{eqnarray}
\begin{eqnarray}\label{Vc}
&&V_{c\bf k}(j+\nu,j,j',j'+\nu)=-\frac{1}{2}f_{2\bf k}(j+\nu,j'+\nu)\nonumber\\
&&\quad\quad\times\bigg\{a_2(j'+\nu,j+\nu)\big[a_1(j)+z_{1\bf k}(j,j')\big]\nonumber\\
&&\quad\quad-\frac{3\delta_{\nu0}}{2(T+\zeta)}a_1(j)a_1(j') \nonumber\\
&&\quad\quad+a_1(j'+\nu)\big[a_2(j,j')+z_{2\bf k}(j,j')\big]\nonumber\\
&&\quad\quad+a_3(j'+\nu,j+\nu)\big[a_4(j,j')+z_{4\bf k}(j,j')\big]\nonumber\\
&&\quad\quad+a_4(j'+\nu,j+\nu)\big[a_3(j,j')+z_{3\bf k}(j,j')\big]\bigg\},
\end{eqnarray}
where
\begin{eqnarray*}
&&f_{1\bf k}(j,j')=\bigg[1+\frac{1}{4}a_1(j)a_1(j'){\cal T}_{\bf k}(j,j')\bigg]^{-1},\\
&&f_{2\bf k}(j,j')=\bigg[1-\frac{3}{4}a_1(j)a_1(j'){\cal T}_{\bf k}(j,j')\bigg]^{-1},\\
&&{\cal T}_{\bf k}(j,j')=N^{-1}\sum_{\bf k'}\theta({\bf k+k'},j)\theta({\bf k'},j'),
\end{eqnarray*}
and quantities $y_{i\bf k}(j,j')$ and $z_{i\bf k}(j,j')$, $i=1,\ldots 4$ are solutions of two following systems of four linear equations:
\begin{eqnarray}\label{ykj}
&&y_{i\bf k}(j,j')=b_{i\bf k}(j,j')+\bigg[c_{i2}({\bf k},j-j')- \frac{\delta_{jj'}}{T+\zeta}\nonumber\\
&&\;\times c_{i1}({\bf k},j-j')\bigg] y_{1\bf k}(j,j')+c_{i1}({\bf k},j-j')y_{2\bf k}(j,j')\nonumber\\
&&\;+c_{i4}({\bf k},j-j')y_{3\bf k}(j,j')+c_{i3}({\bf k},j-j')y_{4\bf k}(j,j'),
\end{eqnarray}
\begin{eqnarray}\label{zkj}
&&z_{i\bf k}(j,j')=d_{i\bf k}(j,j')-e_{i2}({\bf k},j-j')z_{1\bf k}(j,j')\nonumber\\
&&\;-e_{i1}({\bf k},j-j')z_{2\bf k}(j,j')-e_{i4}({\bf k},j-j')z_{3\bf k}(j,j')\nonumber\\
&&\;-e_{i3}({\bf k},j-j')z_{4\bf k}(j,j').
\end{eqnarray}
Thus, the solution of the two Bethe-Salpeter equations was reduced to the solution of two small systems of linear equations (\ref{ykj}) and (\ref{zkj}). The equations depend parametrically on {\bf k}, $j$ and $j'$. In these equations,
\begin{eqnarray*}
&&b_{i\bf k}(j,j')=-\frac{1}{4}a_i(j,j')a_1(j)a_1(j'){\cal T}_{\bf k}(j,j')f_{1\bf k}(j,j')\\
&&\quad+\bigg[a_2(j,j')-\frac{\delta_{jj'}}{T+\zeta}a_1(j)\bigg]c_{i1}({\bf k},j-j')\\
&&\quad+a_1(j)c_{i2}({\bf k},j-j')+a_4(j,j')c_{i3}({\bf k},j-j')\\
&&\quad+a_3(j,j')c_{i4}({\bf k},j-j'),\\
&&d_{i\bf k}(j,j')=\frac{3}{4}a_i(j,j')a_1(j)a_1(j'){\cal T}_{\bf k}(j,j')f_{2\bf k}(j,j')\\
&&\quad-a_2(j,j')e_{i1}({\bf k},j-j')-a_1(j)e_{i2}({\bf k},j-j')\\
&&\quad-a_4(j,j')e_{i3}({\bf k},j-j')-a_3(j,j')e_{i4}({\bf k},j-j'),\\
&&c_{ii'}({\bf k},\nu)=\frac{T}{2}\sum_j a_i(j+\nu,j)a_{i'}(j,j+\nu)\\
&&\quad\times{\cal T}_{\bf k}(j+\nu,j)f_{1\bf k}(j+\nu,j),\\
&&e_{ii'}({\bf k},\nu)=\frac{T}{2}\sum_j a_i(j+\nu,j)a_{i'}(j,j+\nu)\\
&&\quad\times{\cal T}_{\bf k}(j+\nu,j)f_{2\bf k}(j+\nu,j).
\end{eqnarray*}
Vertices $V_s$ and $V_c$ in (\ref{Vs}) and (\ref{Vc}) do not depend on spin variables, which, therefore, were omitted. A more detailed derivation of the above equations can be found in \cite{Sherman18}.

Expressions (\ref{Larkin}), (\ref{K}), (\ref{Vs})--(\ref{zkj}) form a closed set of equations for calculating the one-particle Green's function (\ref{Glt}). This set can be solved by iteration. As a starting function in this iteration the Hubbard-I solution \cite{Hubbard63,Hubbard64} was used. This solution is obtained from the above formulas if the irreducible part is approximated by $C_1$ -- the first term in the right-hand side of (\ref{K}) \cite{Vladimir}. No artificial broadening was introduced in these calculations. Derived one-particle Green's functions and vertices were subsequently used for calculating the spin susceptibility and double oc\-cu\-pan\-cy.

\section{Results and discussion}
\subsection{Transition to long-range antiferromagnetic order}
It is worth noting that the above equations correspond to an infinite crystal since infinite sequences of ladder diagrams were summed without any limitations on site indices. In calculations, integrations over {\bf k} were substituted with summations over some meshes of momenta. Such a discretization is a method of numerical integration, which has nothing to do with the crystal finiteness. Indeed, results obtained with 8$\times$8 and 16$\times$16 meshes are very close. For example, the relative difference of the respective DOSs at a given frequency is less than 0.01.

Equations of the previous section describe the transition to the long-range AF order. The transition manifests itself in the vanishing determinant $\Delta({\bf k},j-j')$ of the system of linear equations (\ref{ykj}), which leads to the divergence of quantities $y_i$, the vertex $V_s$ (\ref{Vs}) and the spin susceptibility (\ref{chi}). For all considered values of $U$ the determinant vanishes first at the AF ordering vector ${\bf k}=(\pi,\pi)$ (the intersite distance is set as the unit of length) and $j=j'$. If in the above equations the parameter $\zeta$ is set to zero, in the 2D half-filled system the transition occurs at a finite temperature $T_{\rm AF}$, which for large $U$ is approximately equal to $0.24t$ and decreases for smaller repulsions \cite{Sherman18}. This finite transition temperature violates the Mermin-Wagner theorem \cite{Mermin} and impedes the attainment of lower temperatures.

\begin{figure}[t]
\centerline{\resizebox{0.95\columnwidth}{!}{\includegraphics{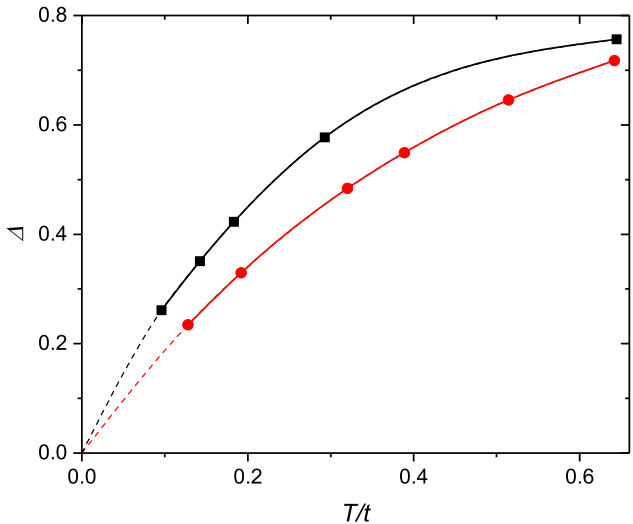}}}
\caption{The temperature dependence of the determinant of the system of linear equations (\protect\ref{ykj}) for half-filling, ${\bf k}=(\pi,\pi)$, $j=j'$ and two values of the repulsion, $U=4t$ (black squares) and $U=8t$ (red circles). Dashed lines show extrapolations to small temperatures. In both cases $\zeta=0.24t$.} \label{Fig2}
\end{figure}As mentioned above, the source of this shortcoming is the terms with the multiplier $1/T$ in the second-order cumulant. The defect can be remedied by adding a small positive quantity $\zeta$ to $T$ in this multiplier. The addend is determined by cumulants of higher orders, which form corrections to the second-order cumulant, the building block of the ladder sequence of $V_s$. Complicated expressions of these corrections do not allow us to calculate $\zeta$ from the serial expansion. Instead, in this work, $\zeta$ is determined from the condition $T_{\rm AF}=0$. The tendency toward the AF ordering is most pronounced at half-filling, which follows from the fact that $\Delta({\bf k}=(\pi,\pi),j-j'=0)$ reaches its minimal value in this case. Therefore, $\zeta$ is determined from the condition $\Delta({\bf k}=(\pi,\pi),j-j'=0)=0$ for $T=0$ and $\mu=U/2$. Doping suppresses the tendency to the AF order due to an increased mobility of electrons. As a consequence, moderate variations of $\zeta$ do not appreciably change $\Delta$. Therefore, in this work, $\zeta$ determined for $\mu=U/2$ is used also in the case of doping. One of the aims of this section is to demonstrate the reasonableness of this approximation. Its validity reveals itself in a satisfactory agreement of calculated properties with Monte Carlo and NLCE data as well as results of experiments with ultracold atoms in 2D optical lattices. The value of $\zeta$ appears to be equal to $0.24t$ for $U\gtrsim 4t$, $0.12t$ for $U=2t$ and $0.05t$ for $U=t$. Figure~\ref{Fig2} demonstrates the temperature behavior of $\Delta({\bf k}=(\pi,\pi),j-j'=0)$ for two values of $U$ and the respective $\zeta$. As seen from the figure, an appropriate $\zeta$ really zeroizes $T_{\rm AF}$.

It is worth noting here that there are other ways for determining $\zeta$. Calculations in the dynamic vertex approximation also face the problem of violating the Mermin-Wagner theorem \cite{Katanin,Rohringer16}. In this approach, the parameter introduced to decrease the magnetic correlation length is determined from the sum rule for the self-energy \cite{Vilk,Katanin} or for susceptibility \cite{Rohringer16}. In that way one can take into account a dependence of this parameter on the electron concentration $\bar{n}=2\langle n_{\bf n\sigma}\rangle$. However, the employment of the sum rule for the self-energy needs analytic continuation to the real frequency axis, which introduces additional inaccuracies. The susceptibility -- the two-particle Green's function -- is also calculated with a lower accuracy than $G({\bf k},j)$. The above-discussed method for estimating $\zeta$ is deprived of these inaccuracies. However, this method does not allow us to determine the dependence of $\zeta$ on $\bar{n}$.

As in \cite{Sherman18}, no indication of a divergence of the charge susceptibility was found in the present cal\-cu\-la\-ti\-ons -- the determinant of the system (\ref{zkj}) varies comparatively we\-a\-k\-ly as the temperature decreases.

\begin{figure}[t]
\centerline{\resizebox{0.95\columnwidth}{!}{\includegraphics{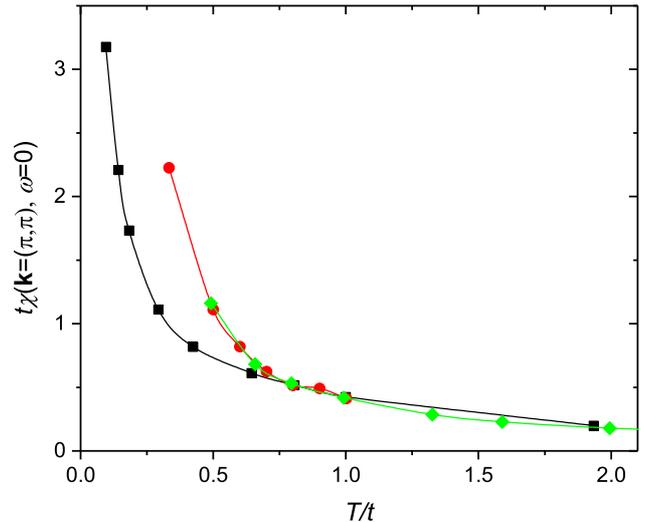}}}
\caption{The temperature variation of the zero-frequency staggered spin susceptibility $\chi({\bf k}=(\pi,\pi),\omega=0)$ at half-filling for $U=4t$. The SCDT results are shown by black squares, red circles and green rhombuses are Monte Carlo data obtained in 4$\times$4 \protect\cite{Bickers} and 6$\times$6 lattices \protect\cite{Hirsch}, respectively.} \label{Fig3}
\end{figure}
Figure~\ref{Fig3} demonstrates the temperature dependence of the zero-frequency staggered susceptibility $\chi({\bf k},\omega)$, ${\bf k}=(\pi,\pi)$, $\omega=0$, calculated for half-filling and $U=4t$. As expected, the quantity diverges as $T\rightarrow 0$. For comparison two dependencies $\chi(T)$ obtained in Monte Carlo simulations \cite{Hirsch,Bickers} are also shown. Lattices used in these simulations get AF ordered when the AF correlation length becomes comparable to their sizes that happens at finite temperatures. As a consequence staggered susceptibilities of finite lattices bend upward at larger $T$ in comparison with the SCDT result. Definitions of spin susceptibilities in various works differ by constant factors. In figure~\ref{Fig3} and below, values of $\chi$ are brought to the definition (\ref{clt}).

\subsection{Uniform susceptibility}
\begin{figure}[t]
\centerline{\resizebox{0.95\columnwidth}{!}{\includegraphics{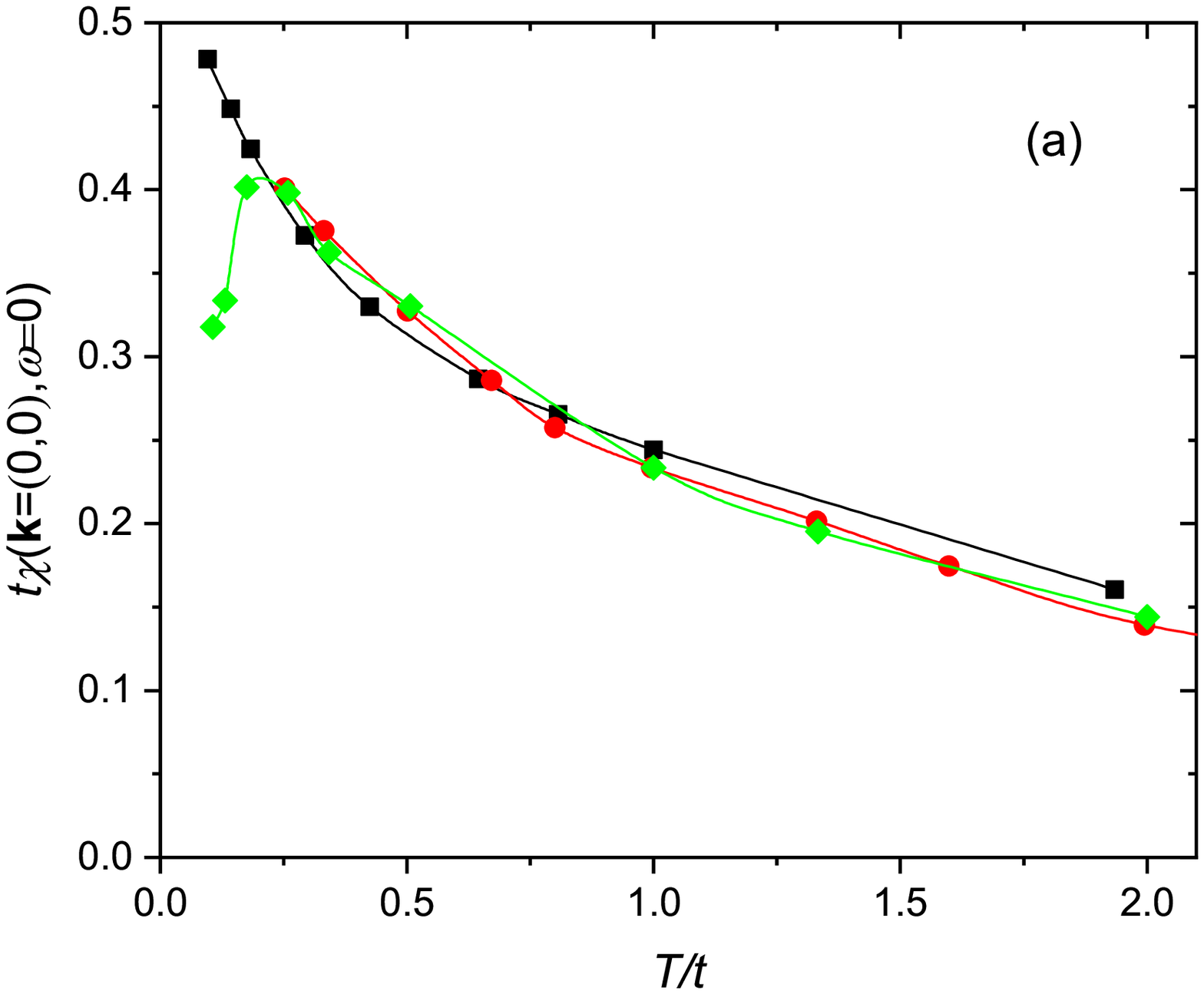}}}
\vspace{1em}
\centerline{\resizebox{0.95\columnwidth}{!}{\includegraphics{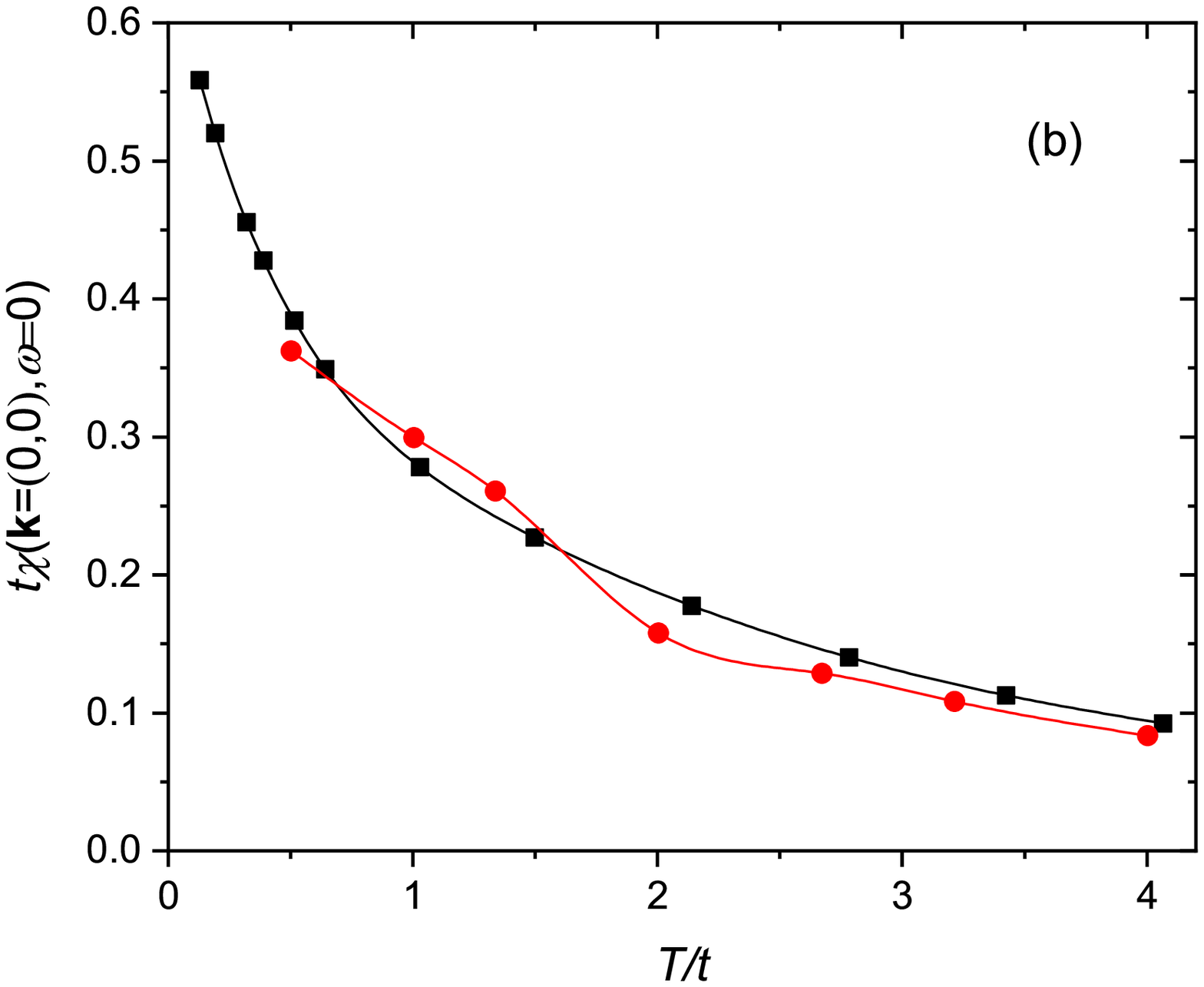}}}
\caption{The temperature variation of the zero-frequency uniform spin susceptibility $\chi({\bf k}=(0,0),\omega=0)$ at half-filling for $U=4t$ (a) and $U=8t$ (b). The SCDT results are shown by black squares, red circles in panel (a) are Monte Carlo data obtained in a 6$\times$6 lattice \protect\cite{Hirsch}, green rhombuses are Monte Carlo results in a 8$\times$8 lattice  \protect\cite{Moreo}, red circles in panel (b) are Monte Carlo data in a 6$\times$6 lattice from \protect\cite{Hirsch83}.} \label{Fig4}
\end{figure}
Figure~\ref{Fig4} shows the temperature dependence of the zero-frequency uniform susceptibility $\chi({\bf k}=(0,0),\omega=0)$ for two values of $U$. Our results are compared with Monte Carlo data \cite{Hirsch,Moreo,Hirsch83} obtained on lattices of different sizes. The agreement of the SCDT re\-sults with these Monte Carlo data is in general sa\-tis\-fac\-to\-ry, except for the region $T\lesssim 0.2t$ in panel (a). Here the Monte Carlo results from \cite{Moreo} demonstrate a steep downturn in the dependence, while our su\-s\-cep\-ti\-bi\-li\-ty continues to grow. A similar non-mo\-no\-to\-no\-us temperature dependence of the uniform su\-s\-cep\-ti\-bi\-li\-ty is observed also in the Heisenberg model \cite{Okabe}. In the Hubbard model, such non-monotonous dependencies were obtained using the dynamic vertex approximation \cite{Li} and the dual fermion approach \cite{Rohringer,Loon}. It is supposed that the difference between these results and ours for small temperatures is connected with a larger set of diagrams used in the mentioned works. In particular, we took into account the simplest ir\-re\-du\-ci\-b\-le four-leg vertex -- the second-order cumulant -- and neglected more complex vertices. This drawback will be corrected in the future.

\begin{figure}[t]
\centerline{\resizebox{0.95\columnwidth}{!}{\includegraphics{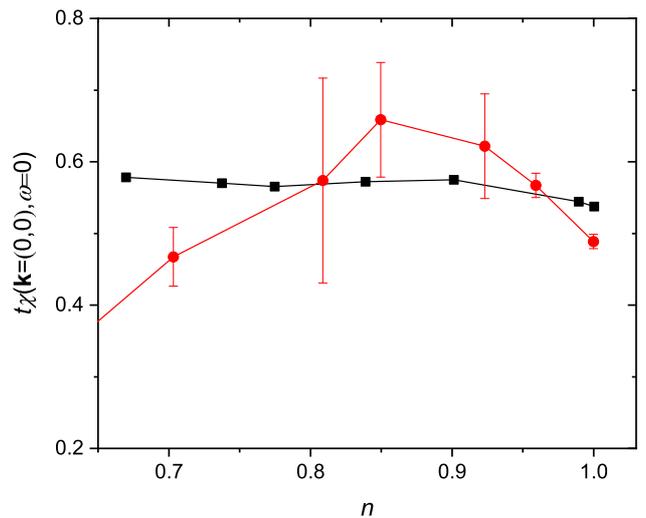}}}
\caption{The zero-frequency uniform spin susceptibility $\chi({\bf k}=(0,0),\omega=0)$ as a function of the electron concentration for $U=10t$ and $T=0.25t$. The SCDT results are shown by black squares, red circles are Monte Carlo data obtained in a 4$\times$4 lattice \protect\cite{Moreo}.} \label{Fig5}
\end{figure}
Figure~\ref{Fig5} demonstrates the zero-frequency uniform spin susceptibility as a function of the electron con\-cen\-tra\-ti\-on for a fixed temperature. For comparison results of Monte Carlo simulations in a 4$\times$4 lattice for the same parameters are also shown. As seen from the figure, the results are consistent with each other to within the errors nearly in all considered range of $\bar{n}$. As in the numeric experiment, in the SCDT results, there is a weak maximum of $\chi$ near $\bar{n}=0.9$. However, after a shallow minimum near $\bar{n}=0.77$ our susceptibility starts to grow with a reduction of the concentration, while in the Monte Carlo simulations it decreases monotonously.

\subsection{Double occupancy, square of site spin and spin structure factor}
The double occupancy $D=\langle n_{\bf l\uparrow}n_{\bf l\downarrow}\rangle$ is an essential quantity characterizing electron correlations. In this work, it is calculated from the relation
\begin{equation}\label{nn}
D=\frac{T}{UN}\sum_{{\bf k}j}\exp(i\omega_j\eta)G({\bf k},j)\Sigma({\bf k},j),\quad \eta\rightarrow+0,
\end{equation}
where
\begin{equation}\label{self-energy}
\Sigma({\bf k},j)=i\omega_j-t_{\bf k}+\mu-G^{-1}({\bf k},j)
\end{equation}
is the self-energy. Equation (\ref{nn}) follows from the equation of motion for Green's function $G({\bf k},j)$ and the Dyson equation \cite{Vilk}.

To obtain the correct result in the summation over $j$ in (\ref{nn}) a large number of terms is necessary. In the iteration procedure described in the previous section, Green's functions with $j\leq 100$ are calculated, which is not enough for the summation. Therefore, for $j>100$ Green's function is approximated by its asymptote \cite{Vilk,Kalashnikov,White}
\begin{equation}\label{asymptote}
G({\bf k},j)=\frac{1}{i\omega_j}+\frac{M_{1\bf k}}{(i\omega_j)^2}+\frac{M_{2\bf k}}{(i\omega_j)^3},
\end{equation}
where
\begin{eqnarray}
M_{1\bf k}&=&t_{\bf k}-\mu+\frac{1}{2}U\bar{n}, \nonumber\\[-1.3ex]
&&\label{moments}\\[-1.3ex]
M_{2\bf k}&=&(t_{\bf k}-\mu)^2+U(t_{\bf k}-\mu)\bar{n}+\frac{1}{2}U^2\bar{n}.\nonumber
\end{eqnarray}
It was found that calculated Green's functions settle into this asymptote at $j\ll 100$.

\begin{figure}[t]
\centerline{\resizebox{0.95\columnwidth}{!}{\includegraphics{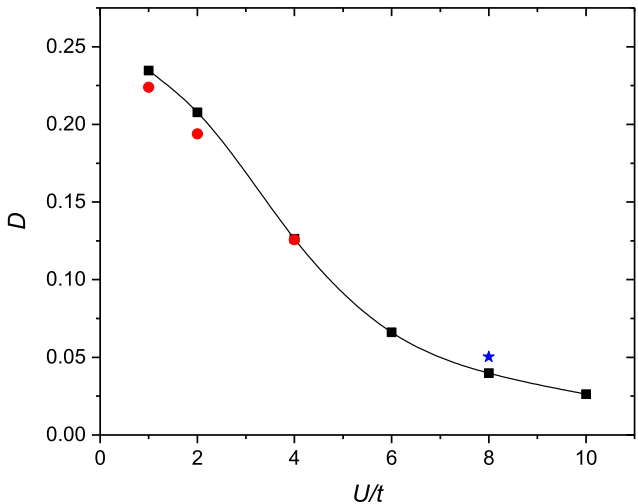}}}
\caption{The double occupancy $D=\langle n_{\bf l\uparrow}n_{\bf l\downarrow}\rangle$ as a function of the Hubbard repulsion. Calculations were carried for half-filling and $T=t/6$. The SCDT results are shown by black squares, red circles are Monte Carlo data obtained in a 12$\times$12 lattice \protect\cite{Moreo90}. The blue star is the result of the dynamic cluster approximation derived for $T=0.25t$ with the extrapolation to the thermodynamic limit \protect\cite{LeBlanc}.} \label{Fig6}
\end{figure}
The double occupancy calculated using the above eq\-u\-a\-ti\-ons is shown in figure~\ref{Fig6}. As expected, it tends to zero for large repulsions and to the uncorrelated limit $1/4$ for small $U$. As seen from the figure, our results are in satisfactory agreement with data of Monte Carlo simulations \cite{Moreo90} and dynamic cluster approximation \cite{LeBlanc}. It should be underlined that the agreement is achieved also for the small repulsion $U=t$. This fact supports our supposition that the SCDT, which was initially devised for the case of strong electron correlations, is also applicable for the case of weak correlations if charge fluctuations are properly included in self-consistent calculations.

\begin{figure}[t]
\centerline{\resizebox{0.95\columnwidth}{!}{\includegraphics{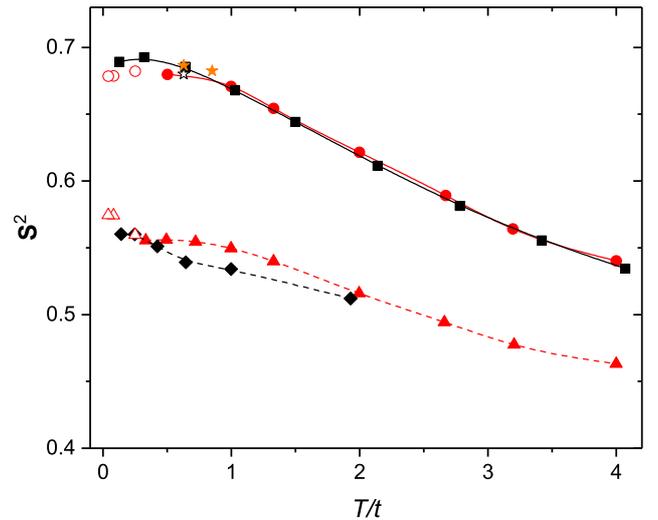}}}
\caption{The temperature dependence of the squared site spin $\langle{\bf S}_{\bf l}^2\rangle$ calculated for half-filling. Black squares and rhombuses are SCDT results for $U=8t$ and $U=4t$, respectively. Red filled circles and triangles are Monte Carlo data obtained for the same repulsions in a 6$\times$6 lattice \protect\cite{Hirsch}. Red open circles and triangles are Monte Carlo results calculated in a 10$\times$10 lattice in \protect\cite{Varney}. Experimental data \protect\cite{Drewes} obtained in a 2D optical lattice are shown by orange stars, the open star is the NLCE result from \protect\cite{Khatami}.} \label{Fig7}
\end{figure}

The double occupancy is connected with the squ\-a\-re of the site spin $\langle{\bf S}^2_{\bf l}\rangle$ by the relation
\begin{equation}\label{S2}
\langle{\bf S}^2_{\bf l}\rangle=\frac{3}{4}\bar{n}-\frac{3}{2}D,
\end{equation}
which follows from the equation $s^z_{\bf l}=(n_{\bf l\uparrow}-n_{\bf l\downarrow})/2$ and the rotation symmetry of spin components. Results of our calculations using (\ref{nn})--(\ref{S2}) are shown in figure~\ref{Fig7}. As might be expected, $\langle{\bf S}^2\rangle$ is larger for smaller $T$ and larger $U$, for which the square of the spin tends to its localized limit $S(S+1)=3/4$ at half-filling. For comparison results of Monte Carlo calculations \cite{Hirsch,Varney} are also shown in figure~\ref{Fig7}. The agreement of our calculated $\langle{\bf S}^2\rangle$ with these results is satisfactory. Of special note is the behavior of $\langle{\bf S}^2\rangle$ at low temperatures, where our results reproduce correctly the different temperature variation of the quantity for $U=8t$ and $4t$ -- its weak reduction with decreasing $T$ in the former case and a perceptible growth with the subsequent flattening in the latter. Our results are also in satisfactory agreement with data of an experiment with ultracold fermionic atoms in a 2D optical lattice \cite{Drewes} and of the NLCE calculations \cite{Khatami}.

\begin{figure}[t]
\centerline{\resizebox{0.95\columnwidth}{!}{\includegraphics{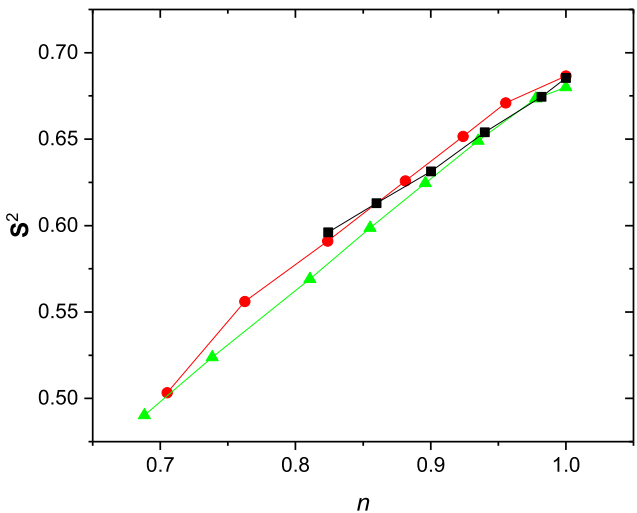}}}
\caption{The concentration dependence of the squared site spin $\langle{\bf S}_{\bf l}^2\rangle$. SCDT results for $U=8t$ and $T=0.64t$ are shown by black squares. Red circles and green triangles are data of experiments in a 2D optical lattice and NLCE calculations, respectively \protect\cite{Drewes}. These data were obtained for $U=8.2t$ and $T=0.63t$.} \label{Fig8}
\end{figure}

The concentration dependence of $\langle{\bf S}^2\rangle$ is shown in fi\-gu\-re~\ref{Fig8}. The range of $\bar{n}$, in which the SCDT calculations were carried out, is limited by the first condition (\ref{condition}). In this range, our results are in satisfactory agreement with the NLCE data and outcomes of experiments in a 2D optical lattice \cite{Drewes}.

\begin{figure}[tbh]
\centerline{\resizebox{0.95\columnwidth}{!}{\includegraphics{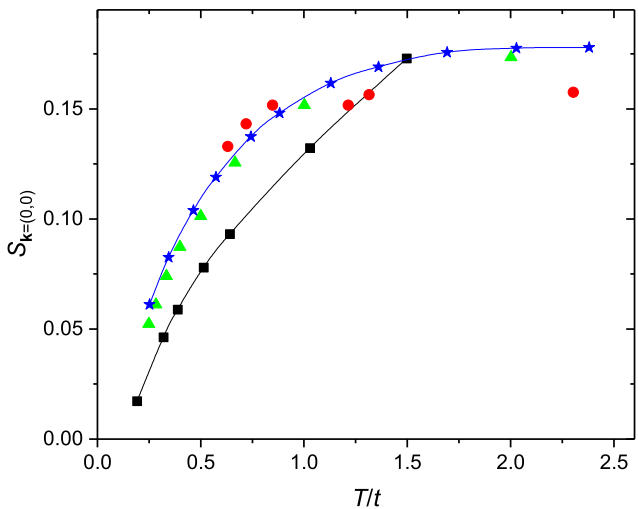}}}
\caption{The temperature dependence of the spin structure factor $S_{{\bf k}=(0,0)}$ at half-filling. Black squares, blue stars, and green triangles are results of SCDT, NLCE \protect\cite{Khatami} and Monte Carlo simulations \protect\cite{Paiva}, respectively. The results are calculated for $U=8t$. Red circles are data of experiments in a 2D optical lattice measured for $U=8.2t$ \protect\cite{Drewes}.} \label{Fig9}
\end{figure}
In figure~\ref{Fig9}, our data on the temperature dependence of the spin structure factor
\begin{equation}\label{ssf}
S_{{\bf k}=(0,0)}=\frac{1}{2}\sum_{\bf l}\langle s^-_{\bf l}s^+_{\bf 0}\rangle=\frac{T}{2}\sum_{\nu}\chi({\bf k}=(0,0),\nu)
\end{equation}
are compared with results of NLCE \cite{Khatami}, Monte Carlo simulations \cite{Paiva} and experiments with ultracold atoms in a 2D optical lattice \cite{Drewes}. Again the range of temperatures, in which the SCDT calculations were performed, was limited by the conditions (\ref{condition}). Our obtained dependence is similar to those derived by other methods. However, the numerical deviation of our results from others is comparatively large. As mentioned above, the two-particle Green's function $\chi({\bf k},\nu)$ is calculated with a lower accuracy than $G({\bf k},j)$. This is the reason for the mentioned deviation. Due to the same reason $\langle{\bf S}^2\rangle$ data calculated directly from $\chi({\bf k},\nu)$ are in a worse agreement with experimental results than those derived from  equations~(\ref{nn})--(\ref{S2}) and shown in figure~\ref{Fig7}. We hope that the mentioned refinement of the four-leg vertex will remedy this flaw.

\subsection{Spin-polaron band}
\begin{figure}[t]
\centerline{\resizebox{0.95\columnwidth}{!}{\includegraphics{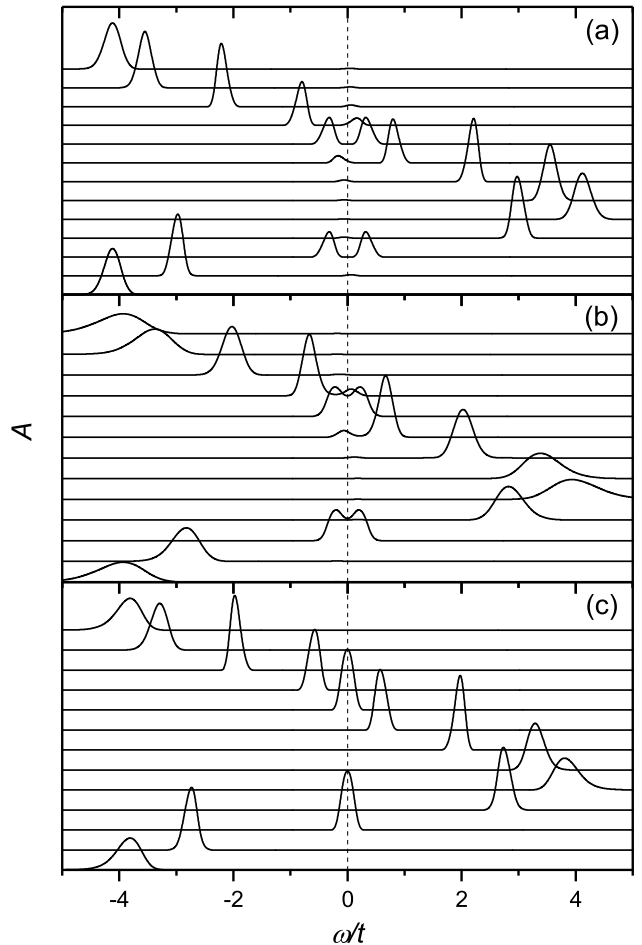}}}
\caption{Spectral functions for momenta along the route $(0,0)$--$(\pi,0)$--$(\pi,\pi)$--$(0,0)$ (from top to bottom) for half-filling, $U=2t$, $T=0.08t$ (a), $0.24t$ (b) and $0.37t$ (c).} \label{Fig10}
\end{figure}
As mentioned above, introducing the parameter $\zeta$ into formulas of the previous section allowed us to attain temperatures, which are much lower than those in the previous SCDT calculations \cite{Sherman18}. In this temperature region, in the case of moderate repulsions, a new feature was revealed in the spectral function $A({\bf k,\omega})=-\pi^{-1}{\rm Im}G({\bf k,\omega})$. This subsection is mainly devoted to a description of this feature. At half-filling, in cases of weak and strong repulsions, no qualitative changes in spectra were found on cooling to these temperatures from those attained in \cite{Sherman18}. Nevertheless, the temperature behaviour of spectra for these repulsions is also con\-si\-de\-red here, since in \cite{Sherman18} this question was mentioned only briefly. In this consideration, the maximum entropy method \cite{Press,Jarrell,Habershon} was used for the analytic continuation of calculated Green's functions from the imaginary to the real axis.

Figure~\ref{Fig10} demonstrates temperature evolution of spectra for the case of small repulsions, $U=2t$. For the low temperature $T=0.08t$ (panel (a)) there are two nonintersecting bands, one of which is located below the Fermi level and the other above it. They approach each other at momenta on the boundary of the magnetic Brillouin zone. In the case of low $T$, there is a narrow gap in the DOS at the Fermi level \cite{Sherman18}, which indicates the separation of the bands. This case corresponds to the Slater regime \cite{Slater} of weak correlations, in which the gap is related to long-range AF fluctuations. With increasing temperature, the bands come closer and closer together (panel (b)) and finally merge into one band crossing the Fermi level (panel (c)).

\begin{figure}[t]
\centerline{\resizebox{0.95\columnwidth}{!}{\includegraphics{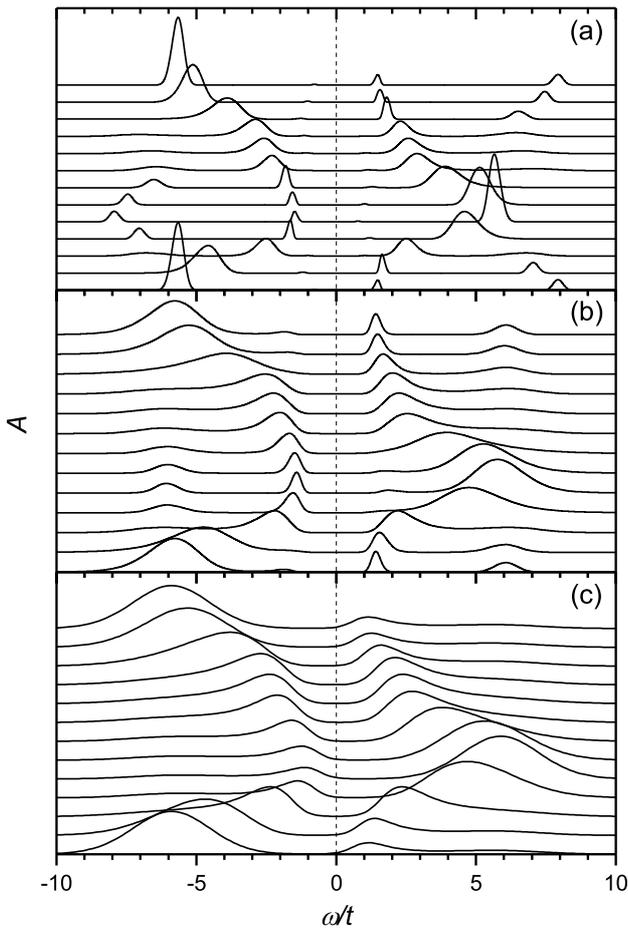}}}
\caption{Same as in figure~\protect\ref{Fig10} but for $U=8t$, $T=0.13t$ (a), $0.32t$ (b) and $0.51t$ (c).} \label{Fig11}
\end{figure}
The opposite case of strong correlations is shown in fi\-gu\-re~\ref{Fig11}. For low temperatures, this case is cha\-rac\-te\-ri\-zed by the four-band structure (panels (a) and (b)) observed first in Monte Carlo simulations \cite{Grober}. As discussed in \cite{Sherman18}, the structure stems from the Mott gap (or, in the doped case, a dip related to it) and two intensity suppressions near frequencies of the Hubbard atom $-\mu$ and $U-\mu$, arising due to a strong reabsorption of electrons on these frequencies. In the doped case, one of these intensity suppressions corresponds to the high-energy anomaly or waterfall observed in photoemission of several families of cup\-ra\-tes \cite{Ronning,Graf,Valla}. As the temperature increases, the four-band structure is gradually transformed into two subbands of the Hubbard spectrum \cite{Hubbard63,Hubbard64} (panel (c)).

\begin{figure}[t]
\centerline{\resizebox{0.94\columnwidth}{!}{\includegraphics{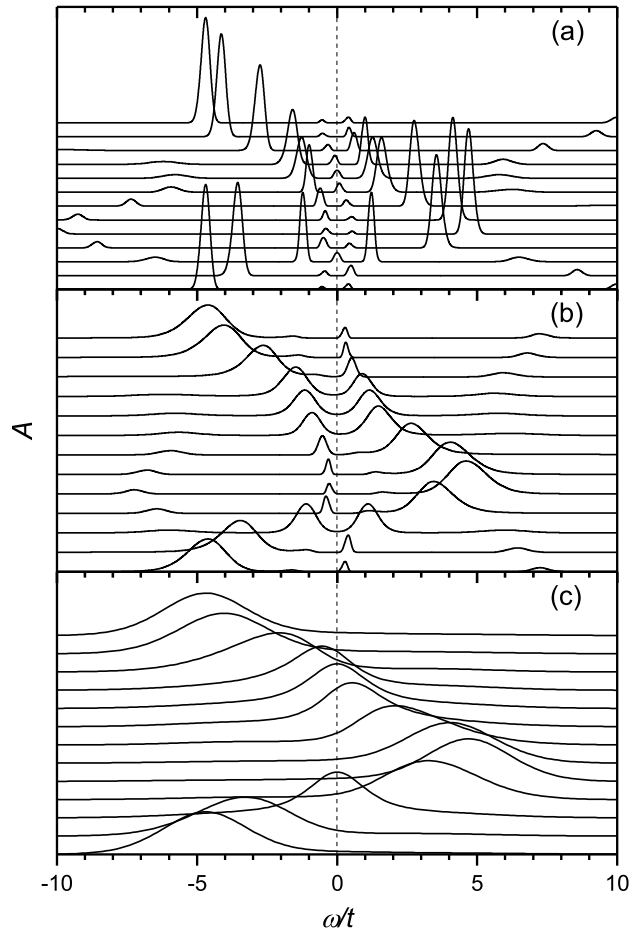}}}
\caption{Same as in figure~\protect\ref{Fig10} but for $U=5.1t$, $T=0.12t$ (a), $0.2t$ (b) and $T=0.82t$ (c).} \label{Fig12}
\end{figure}
\begin{figure}[htb]
\centerline{\resizebox{0.95\columnwidth}{!}{\includegraphics{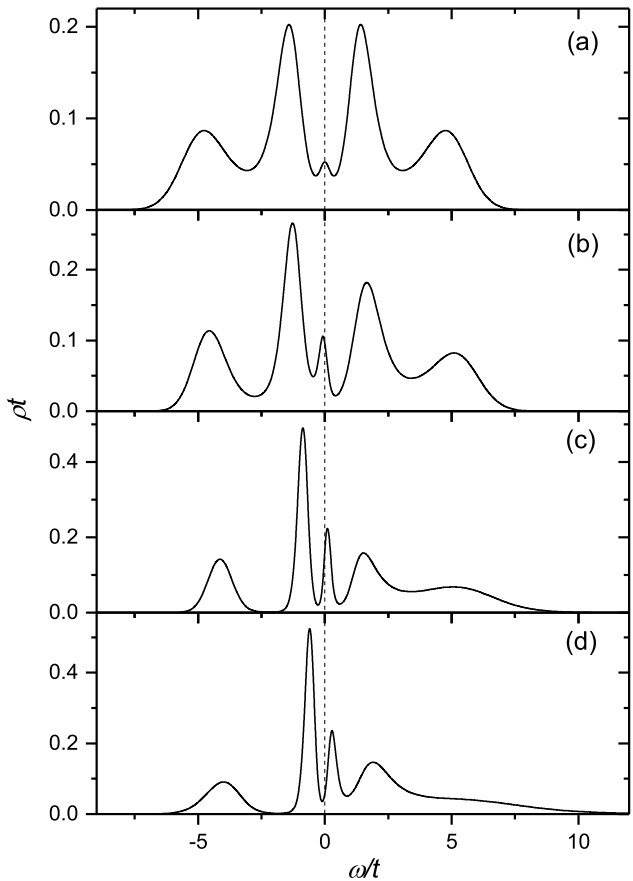}}}
\caption{Densities of states corresponding to $U=5.1t$, $T=0.12t$ and electron concentrations $\bar{n}=1$ (a), 0.98 (b), 0.91 (c) and 0.84 (d).} \label{Fig13}
\end{figure}

The case of intermediate repulsions is shown in fi\-gu\-re~\ref{Fig12}. As seen in panels (a) and (b), spectral functions have some features similar to those observed in both previous cases. In addition, at very low temperatures a new peculiarity reveals itself in spectra -- between two nonintersecting bands resembling those in figure~\ref{Fig10}(a) there is a low-intensity band, which crosses the Fermi level (panel (a)). In the DOS, the band produces a weak peak at the Fermi level on the bottom of the dip inherent in the bad-metal state (see figure~\ref{Fig13}(a)).
The width of this band is of the order of the exchange constant $J=4t^2/U$ that points to its spin-polaron nature -- the band is formed from bound states of electrons and spin excitations. The latter, slower, excitations determine this bandwidth. The band is inherently similar to the spin-polaron band in the $t$-$J$ \cite{Schmitt,Ramsak,Sherman94}. From the comparison of figures~\ref{Fig11}(a) and \ref{Fig12}(a) one can conclude that at half-filling and low temperatures the spin-polaron band appears with closing the Mott gap when $U$ becomes smaller than the critical value $U_c$ of the Mott metal-insulator transition. In accord with the cellular DMFT \cite{Park}, variational cluster approximation \cite{Balzer} and SCDT \cite{Sherman18}, $U_c\approx 6t$. In conformity with the behavior of the band, its related DOS peak is seen on the Fermi level at low temperatures and $U<U_c$ at half-filling. As follows from figure~\ref{Fig13}, with doping, the peak is held near this level in a wide range of $\bar{n}$. On doping and at low temperatures, the peak appears also in the DOS for $U>U_c$, as soon as the Fermi level leaves the Mott gap. These properties of the DOS peak are similar to features of the quasiparticle peak in the DMFT \cite{Georges}. By their nature, these two peaks are also similar. The DMFT quasiparticle peak is a modified Kondo or Abrikosov-Suhl resonance of the Anderson impurity model \cite{Hewson}. The resonance is a manifestation of bound states of {\em free} electrons and {\em localized} spins of this model. In the SCDT, the DOS peak is a manifestation of bound states also. However, in this approach, they are formed by {\em correlated} electrons and {\em mobile} spin excitations. In our opinion, the latter picture is more appropriate for the 2D case, since in this case the momentum dependence of the electron self-energy is strong \cite{Maier,Rohringer,Sherman17} and the consideration based on the local model can provide only qualitative interpretation.

Besides the DMFT, peaks at the Fermi level in DOSs and spectral functions of the Hubbard model were also obtained by other methods, see, e.g., \cite{Bulut,Luchini,Kyung,Izyumov02}.

\begin{figure}[t]
\centerline{\resizebox{0.95\columnwidth}{!}{\includegraphics{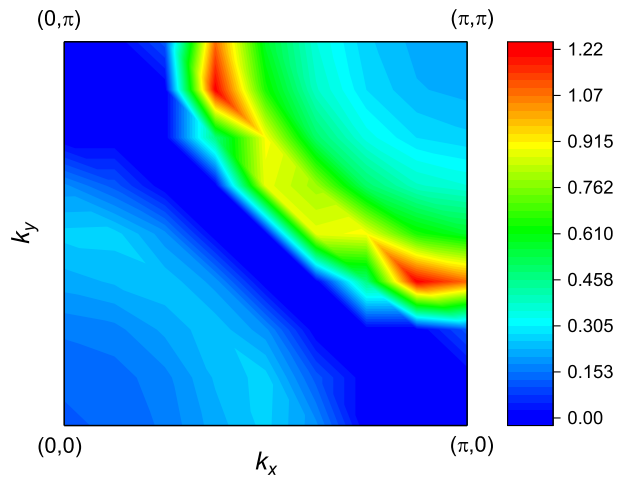}}}
\caption{The contour plot of the spectral function integrated in a narrow frequency window around the Fermi level. $U=5.1t$, $T=0.12t$ and $\bar{n}=0.91$.} \label{Fig14}
\end{figure}

Figure~\ref{Fig14} demonstrates the contour plot of the spectral function integrated in a narrow frequency range near the Fermi level for $U=5.1t$. Contrasting this plot with the analogous figure in \cite{Sherman18} obtained for $U=8t$ we see that for comparable deviations from half-filling Fermi contours are also similar. However, if in the case $U=8t$ the maximum intensity is achieved along the Fermi arc near the nodal point $(\pi/2,\pi/2)$, as in underdoped $p$-type cuprates \cite{Damascelli}, in the case $U=5.1t$ the maximum intensity is located near the boundary of the Brillouin zone.

\section{Conclusion}
In the previous work \cite{Sherman18}, we have found that the SCDT with the used set of diagrams violates the Mermin-Wagner theorem -- an infinite 2D system described by the Hubbard model turns into a state with the long-range AF order at a small but finite temperature $T_{\rm AF}$ for half-filling. In the present work, we eliminated this defect. The correction consists in introducing a parameter into the irreducible four-leg vertex -- the second-order cumulant in the present approximation. The parameter is chosen so that to nullify $T_{\rm AF}$. It appears to be possible for all considered values of the Hubbard repulsion. Equations for the electron Green's function and spin susceptibility were obtained by summing infinite sequences of ladder diagrams and solved by iteration for the ranges of Hubbard repulsions $t\leq U\leq 10t$, temperatures $0.1t\lesssim T\lesssim 4t$ and electron concentrations $0.6\lesssim\bar{n}\leq 1$. We found that temperature and concentration dependencies of the zero frequency uniform susceptibility and the square of the site spin, as well as the variation of the double occupancy with the Hubbard repulsion and the spin structure factor with temperature are in satisfactory agreement with results of Monte Carlo simulations, NLCE and experiments with ultracold atoms in optical lattices. The exception is the uniform susceptibility for half-filling, $U=4t$ and $T<0.2t$. In this case, our calculated quantity grows with de\-cre\-a\-sing $T$, while the Monte Carlo result decreases. It is supposed that the agreement in this region may be improved by extending the set of diagrams taken into consideration. It is worth noting the correct result for the double occupancy obtained for $U$ as small as $t$. This result supports the supposition that the SCDT -- the approach initially invented for the case of strong repulsions -- is able to give quantitatively correct results in the conditions of small repulsions at a proper inclusion of charge fluctuations and self-consistent calculations. The satisfactory agreement of the calculated quantities with results of the mentioned experimental and numerical methods lends support to the validity of the introduced correction.

It allowed us to reach the region of low temperatures and to trace the temperature variations of electron spectra up to this region. At half-filling, for small and large Hubbard repulsions this temperature lowering does not lead to any qualitative changes in comparison with results for higher $T$ in \cite{Sherman18}. For small $U$ and $T$, the spectrum consists of two bands separated by the Slater gap along the boundary of the magnetic Brillouin zone. With increasing temperature, these two bands merge into one band crossing the Fermi level. For large $U$ and small $T$ the spectrum features the four-band structure, which arises due to the Mott gap near the Fermi level and two reabsorption pseudogaps near $\omega=\pm U/2$. With temperature growth, the pseudogaps gradually disappear, and the four-band structure is transformed into two Hubbard subbands. During this heating, a finite intensity appears in the Mott gap due to a temperature broadening of maxima. With the achieved decrease in temperature, qualitative changes in spectral shapes are observed at half-filling for moderate Hubbard repulsions. In this case, for the lowest temperatures attained in \cite{Sherman18}, the electron DOS had a dip near the Fermi level. It turned out that with further cooling there appears a peak at the bottom of this dip. This peak stems from a band with a width of the order of the exchange constant $J=4t^2/U$. This fact points to the spin-polaron nature of the band -- it is formed from bound states of electrons and spin excitations. A similar band exists also in the $t$-$J$ \cite{Schmitt,Ramsak,Sherman94}. Properties of the band and the related DOS peak are similar to those of the DMFT quasiparticle peak and respective states. At half-filling, both peaks are seen near the Fermi level at low temperatures and $U<U_c$. For $U>U_c$ the peaks are observed with doping when the Fermi level leaves the Mott gap. The DMFT peak is a modified Kondo or Abrikosov-Suhl resonance of the Anderson impurity model. This resonance is also a manifestation of bound states. However, in contrast to the SCDT, in which the bound states are formed from correlated electrons and mobile spin excitations, in the Anderson impurity model the states are built up from free electrons and localized spins.

\begin{acknowledgement}
This work was supported by the research project IUT2-27.
\end{acknowledgement}

\end{document}